# Fourier synthesis of radio frequency nanomechanical pulses with different shapes

Florian J. R. Schülein[1,2], Eugenio Zallo[3], Paola Atkinson[3,4], Oliver G. Schmidt[3], Rinaldo Trotta[5], Armando Rastelli[3,5], Achim Wixforth[1,2,6] & Hubert J. Krenner[1,2,6,*]

(1)     Lehrstuhl für Experimentalphysik 1 and Augsburg Centre for Innovative Technologies (ACIT), Universität Augsburg, Universitätsstr. 1, 86159 Augsburg, Germany

(2)     Nanosystems Initiative Munich (NIM), Schellingstraße 4, 80339 München, Germany

(3)     Institute for Integrative Nanosciences, IFW Dresden, Helmholtzstraße 20, 01069 Dresden, Germany

(4)     Institut des NanoSciences de Paris, Sorbonne Universités, UPMC Univ. Paris 06, CNRS UMR7588, 4 Place Jussieu, F-75005 Paris, France.

(5)     Institute of Semiconductor and Solid State Physics, Johannes Kepler Universität Linz, Altenbergerstr. 69, 4040 Linz, Austria

(6)     Center for Nanoscience (CeNS), Ludwig-Maximilians-Universität München, Geschwister-Scholl-Platz 1, 80539 München, Germany

*       hubert.krenner@physik.uni-augsburg.de

**The concept of Fourier synthesis[1] is heavily employed in both consumer electronic products[2] and fundamental research[3]. In the latter, pulse shaping is key to dynamically initialize, probe and manipulate the state of classical or quantum systems. In nuclear magnetic resonance, for instance, shaped pulses have a long-standing tradition[4] and the underlying fundamental concepts have subsequently been successfully extended to optical frequencies[3,5] and even to implement quantum gate operations[6]. Transferring these paradigms to nanomechanical systems requires tailored nanomechanical waveforms. Here, we report on an additive Fourier synthesizer for nanomechanical waveforms based on monochromatic surface acoustic waves. As a proof of concept, we electrically synthesize four different elementary nanomechanical waveforms from a fundamental surface acoustic wave at $f_1 \approx 150$ MHz using a superposition of up to three discrete harmonics fn. We employ these shaped pulses to interact with an individual sensor quantum dot and detect their deliberately and temporally modulated strain component via the opto-mechanical quantum dot response[7,8,9]. Importantly, and in contrast to the direct mechanical actuation by bulk piezoactuators[7], surface acoustic waves provide much higher frequencies (> 20 GHz[10]) to resonantly drive mechanical motion[11]. Thus, our technique uniquely allows coherent mechanical control[12] of localized vibronic modes of optomechanical crystals[13,14], even in the quantum limit when cooled to the vibrational ground state[15].**

In nanomechanics the displacement ($u$), stress ($T$) or strain ($S$) are the defining parameters $A(t)$ of a time ($t$)-dependent nanomechanical waveform (NMWF). To perform additive Fourier synthesis of a desired waveform $A(t)$, a coherent and monochromatic nanomechanical wave of fundamental frequency $f_1$ and a combination of its harmonics $f_n = n \cdot f_1$, $n$ being an integer, are superimposed. Amongst the wealth of the waveforms and impulses which can be generated this way, the square, the sawtooth and the δ-pulse (or impulse) waveforms are elementary: the square waveform allows for fast digital switching between two levels, the sawtooth combines a slow and a fast transient for adiabatic/non-adiabatic switching and the δ-pulse represents a short "kicking" impulse. Their respective Fourier series representations are given by the following analytical expressions:

$$A_{square}(t) = \sum_{n=1,3,5,...} n^{-1} \sin 2\pi n f_1 t$$
$$A_{sawtooth}(t) = \sum_{n=1,2,3,...} n^{-1} \sin 2\pi n f_1 t \quad (1)$$
$$A_{\delta-pulse}(t) = \sum_{n=1,2,3,...} \cos 2\pi n f_1 t.$$

Their monochromatic and coherent nature makes surface acoustic waves (SAWs) ideally suited for this purpose. Moreover, these coherent acoustic phonon modes can be all-electrically excited at precisely defined frequencies[16] and detected in the quantum mechanical single phonon limit[17]. Since their propagation is governed by the speed of sound, SAW wavelengths in the μm and sub-μm range cover radio frequencies (rf) from several tens of megahertz up to the gigahertz range. Such high frequencies are typically inaccessible by other non-resonant, broadband excitation schemes, in particular bulk piezoelements. Apart from their importance in the field of rf signal processing, monochromatic SAWs have been employed over the past decades in a number of fundamental experiments to probe and control transport of charge and spin carriers or excitons in low-dimensional semiconductor structures[18-21], manipulate the occupancy and quantum states in quantum dots (QDs)[22-26], dynamically tune the optical modes in micro- and nanophotonic resonators[27,28] or to coherently control nanomechanical systems[11,12].

Our experimental implementation of a SAW Fourier synthesizer is depicted schematically in Figure 1a. It is based on a piezoelectric SAW chip onto which a delay line is formed by two opposing interdigital transducers (IDT). Employing up to four rf generators and specially designed SAW transducers, we manage to excite the individual Fourier components of a NMWF at IDT1. The pulse is then received by IDT2 and detected by an oscilloscope. This signal is used to actively stabilize the amplitudes and phases of the signal generator outputs via a proportional-integral-derivative (PID) feedback loop. In order to generate the different required harmonics with sufficient efficiency by a single IDT, we apply two advanced transducer geometries derived from inverse Fourier design. These two designs, to which we refer to as Split-4 and Split-52 in the following, are shown in the insets of Figure 1b and 1c, respectively. For both layouts, the longest periodicity determines the acoustic wavelength $\Lambda_1$ of the fundamental resonance. Its frequency is given by $f_1 = c_s/\Lambda_1$, with $c_s = 2860 \frac{m}{s}$ being the phase velocity of a Rayleigh-SAW on GaAs. For the Split-4 design, efficient excitation of *only odd* harmonics $n = 1,3,5$ is possible, while even harmonics are strongly suppressed. In contrast, a Split-52 IDT is optimized for the excitation of a combination of both *even and* *odd* harmonics, *n* = 1,2,3,4,6 (*n* = 5 is suppressed). To demonstrate the nanomechanical function of our synthesizer, we monolithically integrated delay lines of both types on GaAs substrates, which contained a single layer of strain-free GaAs/AlGaAs sensor QDs grown by molecular beam epitaxy[29]. These embedded QDs strongly interact with and thus sense the NMWF's strain field at a depth of *z* = -152 nm below the surface by quantum- and optomechanical coupling[9].

To check the rf characteristics of the fabricated delay line, we recorded the room temperature transmission $S_{21}$ in the frequency range 0 < *f* < 1 GHz, set by the 3 dB bandwidths of the rf components used. The measured spectra are compared in Figure 1b and c for the Split-4 and Split-52 delay lines, respectively. In these data, the corresponding fundamental frequencies are clearly resolved at $f_1^{(4)} = 180$ MHz for the Split-4 and $f_1^{(52)} = 144$ MHz for the Split-52, as expected for the lithographically defined design wavelengths of $\Lambda_1^{(4)} = 16\ \mu m$ and $\Lambda_1^{(52)} = 20\ \mu m$. These fundamentals were chosen to be compatible with the decay time of the QD emission of ~ 1 ns (see Supplementary information). Thus, we can readily probe the dynamic interaction between the NMWF and the QD by time-integrated detection of the stroboscopically excited QD emission. In the rf transmission for both types of IDTs in Figure 1b and c, we detect $f_1$ together with clear signatures of the aforementioned expected higher harmonics (marked by arrows) while the suppressed harmonics are absent.

Both types of IDTs enable us to perform additive Fourier synthesis of NMWFs as defined by equation (1). In the following, we demonstrate the dynamic spectral modulation of our sensor QD by four prototypical waveforms. For the demonstration of this fundamental procedure, we first derive the response of our local sensor to the acoustic field. The mechanical displacement of the Rayleigh-type SAWs has a longitudinal, $u_x$, and a transverse, $u_z$, component. When a positive (negative) electrical potential, $\Phi$, is applied to the IDT, the piezomechanical coupling induces a negative (positive) $u_z \propto -\Phi$. In contrast, the QD's spectral response does not directly follow the displacements $u_x$ and $u_z$, but is given for moderate acoustic amplitudes by the deformation potential (DP) coupling[9,30]. This type of coupling is confirmed by its characteristic dependence on the acoustic amplitude[31], as presented in the Supplementary information. The DP coupling is proportional to the local hydrostatic pressure

$$p = -E_Y \left( \frac{\partial u_x}{\partial x} + \frac{\partial u_z}{\partial z} \right), (2)$$

with $E_Y$ denoting the bulk modulus. Thus, any time-dependent optomechanical response of the sensor QD can be programmed directly by applying a tailored rf signal $V_{\text{IDT}}(t) = \Phi(t)$ to the IDT which induces the required *p* at the position of the QD. In Figure 2, we present numerical calculations for the implementation of four prototypical waveforms: a single frequency sinusoidal oscillation (a), a square wave (b), a sawtooth wave (c) and a δ-pulse (d). The latter are realized by additive synthesis of the fundamental and three harmonics. The upper panels summarize the full mechanical profile of the NMWFs as a function of time *t*. The electric potential, $\Phi$, induced by the piezomechanical coupling is superimposed as colour code and the arrows indicate the resulting gyrating electric field $E = -\nabla\Phi$. The amplitudes of $\Phi(t)$ (dashed red line) and $u_z(t)$ (full black line) evaluated at the depth of the QD are plotted in the centre panels and the induced hydrostatic pressure in the lower panels. The latter resembles exactly the desired four prototypical waveforms given by Equation (1). As we apply the such derived $V_{IDT}(t) = \Phi(t)$ to the IDT, the hydrostatic pressure $p(t)$ of the launched NMWF is expected to programme the desired spectral response of the sensor QD.

We synthesized these four prototypical waveforms in our experiments employing the two IDT layouts presented in Figure 1. For sine and square NMWFs, a Split-4 IDT was used, while the sawtooth and δ-pulse NMWFs were synthesized using a Split-52 IDT. The experimentally accessible bandwidth of 1 GHz allowed Fourier synthesis of the square wave from three ($f_1$, $f_3$, $f_5$) components and of the sawtooth and δ-pulse from four components ($f_1 - f_4$). To experimentally prove the synthesized waveforms, we present in Figure 3 the time-integrated emission spectra of representative single sensor QD emission lines recorded under stroboscopic excitation for the four prototypical NMWFs. The detected intensity is encoded in false colour and plotted in the upper panels as a function of the temporal delay of the excitation laser pulse with respect to the NMWF. Clearly, all programmed NMWFs, sine wave (Figure 3a), square wave (Figure 3b), sawtooth wave (Figure 3c) and δ-pulse wave (Figure 3d), are well reproduced in the spectral response of the sensor QD. Sine and square wave exhibit the expected period $1/f_1^{(4)} = 1/182.7\text{ MHz} = 5.47$ ns of the Split-4 IDT, while sawtooth and δ-pulse are modulated on a timescale set by the fundamental $1/f_1^{(52)} = 1/146\text{ MHz} = 6.85$ ns of the Split-52 IDT. Note that these periods are reduced by ~0.1 ns when compared to the room temperature characterisation in Figure 1. This is as expected due to the increase in sound velocity of the material at low temperatures. A detailed analysis of this data is presented in the lower panels of Figure 3. The observed spectral shifts (symbols) are plotted together with the expected spectral modulation induced by the NMWF's hydrostatic pressure $p(t)$ of the NMWFs (red lines) given by the Fourier series of equation (1). In particular, the observed spectral modulation amplitude of $\delta E_{max} = 150\pm10$ ($100\pm10$) μeV for the sine and square wave (sawtooth and δ-pulse) correspond to a local hydrostatic pressure of $p_{max} = 1.0\pm0.2$ ($0.7\pm0.15$) MPa. Using the linear dependence of $u_z$ and $p$ we extract the optomechanical coupling parameter at the sample surface given by

$$\gamma_{om} = \frac{\partial(\delta E_{max})}{\partial u_z}\bigg|_{z=0}. \quad (3)$$

From our numerical data we obtain $\gamma_{om} = 1800\pm400$ μeV/nm (sine), $3300\pm350$ μeV/nm (square), $2600\pm500$ μeV/nm (sawtooth, fast transient) and $3600\pm600$ μeV/nm (δ-pulse). Details of the applied procedure are laid out in the supplementary information. The stated uncertainties reflect the variations of measured $\delta E_{max}$ of all emission lines detected from the sub-ensemble of QDs located in the laser spot. As expected, the sine, square and sawtooth wave exhibit within the experimental uncertainty similar values, due to the ∝ $n^{-1}$ weighting of the higher Fourier components. In contrast, for the δ-pulse, all components contribute with equal weight (∝ $n$) which manifests itself in the increase of the optomechanical coupling parameter. All these values exceed that of state of the art freestanding and resonantly excited nanowire architecture[7,32] by more than one order of magnitude and can be enhanced substantially by increasing $f_1$. Moreover, we extract the peak-peak transient time of the QD's spectral response [shown as vertical lines marked by arrows in Figure 3]. While for the sine wave the modulation is solely governed by the fundamental, the square wave exhibits fast edges with transient times of 750 ps. For the sawtooth waveform we extract slow and fast transient times of 5.5 ns and 1.35 ns, respectively and for the δ-pulse a temporal width of 850 ps. The experimentally observed times are well reproduced by the spectral response expected for the dynamic pressure evolution of the ideal NMWF.

Finally, we performed a full Fourier analysis in a sine basis of the three experimentally observed NMWFs (for details see Supplementary information). The Fourier coefficients extracted from the QD emission and the phase of the harmonics relative to the fundamental frequency are plotted as coloured bars for the harmonics $n = 2…5$ in Figure 4a and b, respectively. When comparing the experimentally obtained coefficients and phases to the ideal values (grey bars), we find that all excited components are clearly present in the data, while non-excited harmonics are clearly suppressed. The deviations of the Fourier components of higher harmonics, in particular for the δ-pulse waveform, could arise from a small Stark-shift induced by the SAW's piezoelectric field[31] or the limited time resolution due to the time-integrated detection scheme employed[28]. The latter is indeed more dominant for the δ-pulse, since all components contribute with equal strength. This limitation does not apply to the relative phase, which is nicely reproduced in the Fourier analysis. Taken together, our Fourier analysis clearly proves the successful synthesis of a well-defined NMWF in the rf domain.

In conclusion, we implemented a scheme to excite well-defined NMWFs in the rf domain by additive Fourier synthesis of monochromatic SAWs. The developed procedure was applied to generate a desired dynamic pressure field $p(t)$ using an electrical input signal $\Phi(t)$. Such programmed spectral response of a single sensor QD is directly reproduced in the experimental data for the prototypical sinusoidal, square, sawtooth and delta function waveforms. The excellent agreement achieved demonstrates that our procedure can be readily applied to pinpoint tailored rf nanomechanical fields on a sub-nanosecond timescale and transduce these into an optical signal. While our particular scheme requires a tailored pressure field (or stress $T$) given by Equation (2), the electrical input signal can be readily adjusted to set any other tuning parameter. Our technique to synthesize NMWFs up to GHz frequencies is fully compatible with the planar geometry of optomechanical or phoXonic crystals and can be directly applied for acousto-mechanically driven coherent control of these systems in their quantum ground state. Due to the strong optomechanical coupling, Fourier-synthesized acoustic pulses are thus expected to allow for real-time control of quantum coupling realizing a Landau-Zener-based entangling quantum gates[33]. We have shown that due to their very narrow homogeneous linewidth and small footprint, single QDs can be used as local nanomechanical spectrum analysers and the optomechanical interactions can be controlled in the resolved sideband regime[8]. Furthermore, when operating in the resolved sideband regime, optically driven cooling or heating[8] has the potential to enable the coherent manipulation of selected Fourier components down to the single phonon limit. These highly coherent interactions are further underpinned by recent experiments demonstrating coherent coupling between a superconducting qubit and single acoustic quanta[34].

**Methods:**

*Heterostructure and SAW delay lines*

The strain-free GaAs/AlGaAs QDs were grown using molecular beam epitaxy (MBE) on semi-insulating [001]-GaAs substrates using a local droplet etching technique[29]. The QDs are formed by a 3.5 nm layer of GaAs on top of a 7 nm thick $Al_{0.44}Ga_{0.56}As$ layer patterned with Ga-droplet etched nanoholes ($d_{QD} \approx 80 \pm 20$ nm) with a areal density of $\sim 3$ μm$^{-2}$. The QDs are overgrown by a sequence of 112nm $Al_{0.33}Ga_{0.67}As$, 20 nm $Al_{0.44}Ga_{0.56}As$ and a 20 nm GaAs capping layer. Typical emission spectra of these QDs and the measured PL decay time $\tau_{PL} \approx 1$ns are summarized in the Supplementary Information.

Split-4 (31 periods) and Split-52 IDTs (25 periods) with an aperture of $A_{IDT}$ = 400 μm were monolithically integrated on the as-grown GaAs substrate using standard electron beam lithography and a lift-off process (metallization Ti 5 nm, Al 50 nm). The two delay lines were [1$\bar{1}$0]-propagating and 3 mm long for the Split-4 IDTs and [110]-propagating and 3.5 mm long for the Split-52 IDTs.

*Generation of rf signals for NMWF excitation*

The output of up to four rf signal generators is combined using a 4-port power combiner (Mini Circuits ZFSC-4-1+, 3dB bandwidth 1 GHz). The power combiner's output is amplified (Mini Circuits ZHL-42W, 3dB bandwidth 4.2 GHz) and connected to the exciting IDT. The input rf signal and signal detected by the receiving IDT are analysed using a rf oscilloscope (3dB bandwidth 2 GHz). Phase and amplitudes of both signals are analysed in a sine basis to actively stabilize the NMWF using a PID feedback loop.

*Optical spectroscopy*

For stroboscopic μ-PL spectroscopy[26,28,31] samples are placed in a Helium-flow cryostat equipped with custom built integrated rf connections and cooled to $T$ = 10 K. An externally triggered diode laser emitting $\tau_{laser}$~90 ps pulses at a wavelength of 660 nm is focused to a $d_{laser}$ = 1.5 μm spot (probing on average $\sim$ 5 QDs) using a NIR 50× microscope objective to photoexcite carriers in the semiconductor. Stroboscopic excitation is realized by actively phase-locking the train of laser pulses to the rf signals exciting the NMWF. The emission of single QDs is dispersed in a 0.5 m imaging grating monochromator and its time-averaged intensity is detected by liquid $N_2$-cooled Silicon charge coupled device. Due to the finite dispersion over the wide frequency range and frequency dependent attenuation, the transmitted waveform was analysed and stabilised and we restricted our experiments to QDs located at distances < 200 μm from the receiving IDT2. These QDs were positioned well within the SAW soundpath, where effects of lateral dispersion can be readily neglected.

*Time resolution*

The time resolution is on one hand side given by that of the stroboscopic excitation $\tau_{strob}$. Upper and lower boundaries are given by the SAW time delay across the dimensions of the QD and the laser spot and the duration of a laser pulse $d_{QD}/c_s \approx 30\text{ps} < \tau_{laser} < \tau_{strob} < d_{laser}/c_s \approx 500\text{ps}$. Since $\tau_{PL} < 5/f_1^{(4,52)}$, our time-integrated detection effectively averages over only less than ~20% of the fundamental period of the NMWF. This in turn allows us to experimentally resolve the fast temporal features induced by higher harmonics even without sophisticated time-resolved detection schemes[26,28,31].

**Acknowledgements** We gratefully acknowledge support by the Deutsche Forschungsgemeinschaft (DFG) via the Emmy Noether Programme (KR 3790/2-1), the Nanosystems Initiative Munich (NIM), and Sonderforschungsbereich SFB 631, by BMBF via project QuaHL-Rep (Contracts No.01BQ1032 and 01BQ1034) and by the European Union via Seventh Framework Programme 209 (FP7/2007-2013) under Grant Agreement No. 601126 210 (HANAS). We acknowledge enlightening discussions with John H. Davies, Gabriel Bester and Jairo Ricardo Cardenas on strain tuning of quantum dots.

**Author contributions** F.J.R.S., A.W. and H.J.K. designed research. F.J.R.S. designed device, carried out the experiments and performed numerical modelling. F.J.R.S. and H.J.K. performed data analysis and modelling with input from A.W., A.R. and R.T.. E.Z., P.A., A.R. and O.G.S. performed crystal growth and sample characterization. All authors discussed the results. H.J.K., A.W. and F.J.R.S. wrote the manuscript with input from all other authors. H.J.K. supervised the project.

**Additional information**

Supplementary information accompanies this paper at www.nature.com/naturenanotechnology.

Reprints and permission information is available online at http://npg.nature.com/reprintsandpermissions/.

Correspondence and requests for materials should be addressed to H.J.K.

**Figures:**

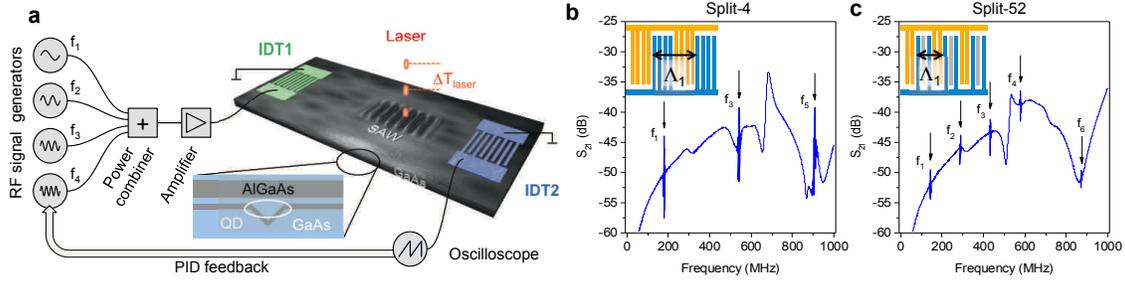

**Figure 1** – Nanomechanical waveform synthesizer – **a** Computer rendered schematic of experimental setup consisting of a GaAs patterned with a SAW delay line. The outputs of up to four RF signal generators are combined, amplified and connected to IDT1 to generated NMWFs. The transmitted NMWF is detected at IDT2 by an oscilloscope. The received signal is used to actively stabilize the excitation RF signal using a PID feedback loop. **b**+**c** RF transmission characteristics ($S_{21}$) of two delay lines of Split-4 (**b**) and Split-52 IDTs (**c**) demonstrating the transmission of odd ($f_1$, $f_3$, $f_5$) and both even and odd harmonics ($f_1$, $f_2$, $f_3$, $f_4$, $f_6$) of the fundamental for Split-4 and Split-52 IDTs, respectively. The insets depict the geometries of the two types of IDTs and the fundamental acoustic wavelength ($\Lambda_1$).

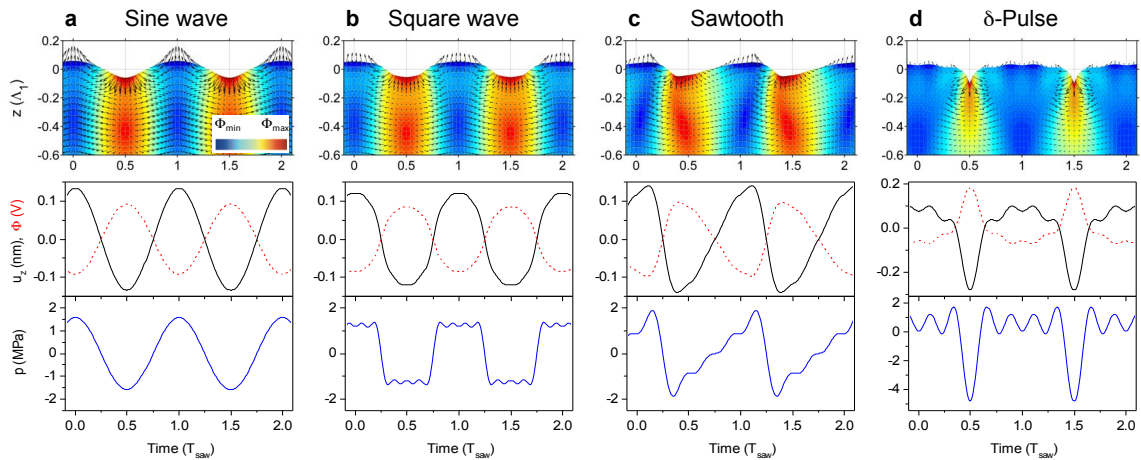

**Figure 2** – Calculated prototypical nanomechanical waveforms of a Sine wave (**a**), Square wave (**b**), Sawtooth wave (**c**) and δ-Pulse (**d**) – *Upper panels:* Calculated mechanical waveform, its associated electric potential (colour coded) and electric field (arrows). Colour scale of electric potential ranges from $\Phi_{min}$=-0.1 V to $\Phi_{max}$= +0.1 V for **a-c** and $\Phi_{max}$= +0.2 V for **d**. *Centre panels:* Extracted vertical displacement component ($u_z$, black line) and electrical potential ($\Phi$, red dashed line) at the position of the sensor QD. *Lower panels:* Corresponding hydrostatic pressure ($p$) given by equation (2).

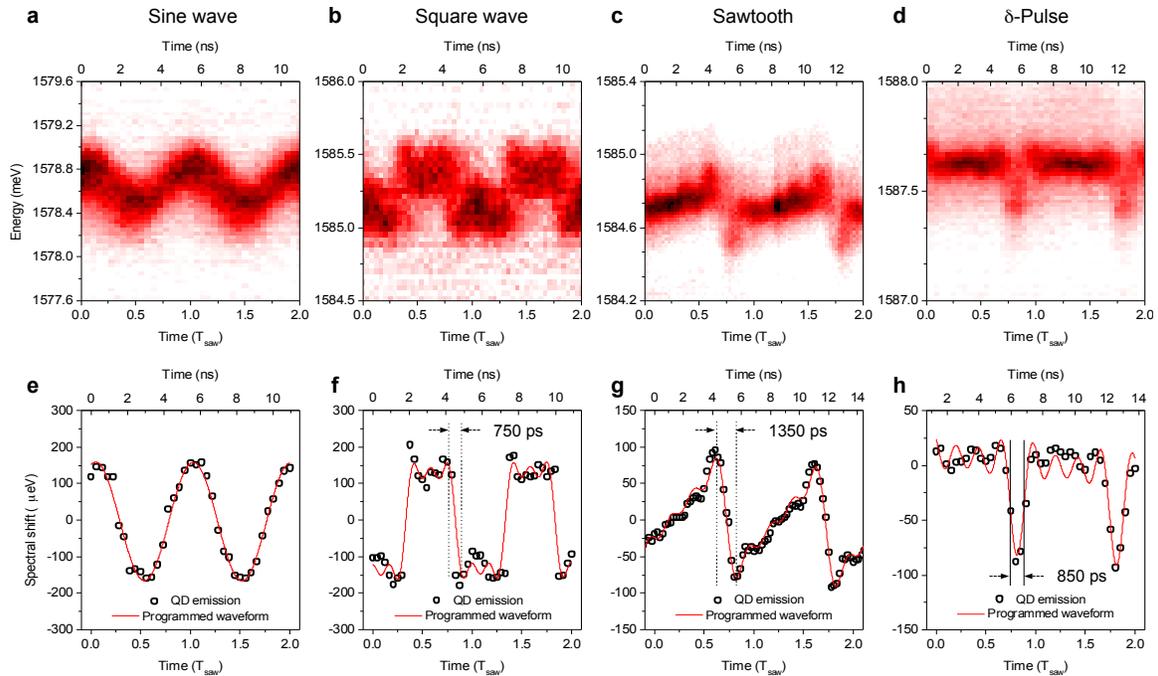

**Figure 3** – Single Quantum Dot sensing of nanomechanical waveforms – *Upper panels:* Stroboscopic PL spectrum of a single sensor QD due to optomechanical coupling to a Sine wave (**a**), Square wave (**b**), Sawtooth wave (**c**) and δ-Pulse (**d**) recorded over two acoustic cycles. Lower panels **e-h** Extracted experimentally observed energy shifts (symbols), achieved switching time and programmed spectral modulations (red line) for all four elementary NMWFs.

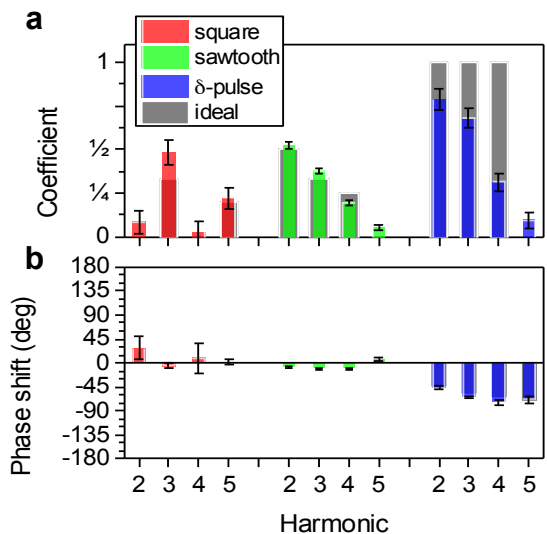

**Figure 4** – Fourier analysis – Comparison between the experimentally observed (coloured bars) and as-programmed (grey bars) Fourier coefficients (**a**) and relative phase with respect to the fundamental (**b**) for the different harmonics for the three non-trivial elementary NMWFs.

# Supplementary Information for "Fourier synthesis radio frequency nanomechanical pulses with different shapes"

Florian J. R. Schülein, Eugenio Zallo, Paola Atkinson, Oliver G. Schmidt, Rinaldo Trotta, Armando Rastelli, Achim Wixforth & Hubert J. Krenner

1. *Quantum Dot characterization*

    a. Time-integrated optical characterization
    In Supplementary Fig. 1, we present an overview PL spectrum. It consists of the dominant emission of the bulk-like GaAs layers (~1.52eV) and the two-dimensional QW of the GaAs layer sandwiched between two AlGaAs barriers at 1.67 eV. The emission of the strain-free QD [1] breaks up into two groups of lines centered at ~1.58 eV and ~1.62eV. For our experiments we restricted to emission lines in the 1.58 eV emission band.

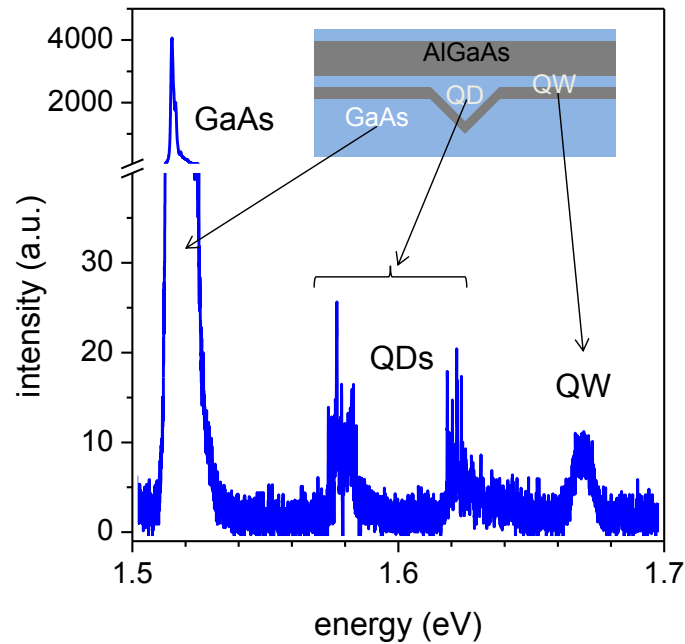

Supplementary Figure 1 – Overview PL spectrum showing emission of the bulk GaAs and the GaAs QW and QDs.

    b. Time-resolved spectroscopy
    In Supplementary Fig. 2, we present a typical PL transient recorded of a single QD emission line (solid blue line). After deconvolution of the instrument response function, IRF (solid black line), we extract a PL decay time of $\tau_{PL} = 1.1$ ns for this particular emission line. The average $\tau_{PL} \approx 1$ns set the

effective integration for a given stroboscopic time delay. Since the fundamental period of our NMWFs is at least a factor of 5 longer, we are able to resolve fast transients in the time evolution of the QD emission. These can only be detected for stroboscopic excitation conditions for which these switchings occur with ~ 1 ns after photoexcitation. Thus the fast rising and falling edges are clearly resolved since they occur *once* per fundamental cycle, while faster ringings, e.g. during the "off"-time of the δ-pulse lead to an overall spectral broadening. A more detailed investigation could be performed in future experiments employing a combination of stroboscopic excitation and time-resolved detection [2] or high-resolution optical spectral analysis [3].

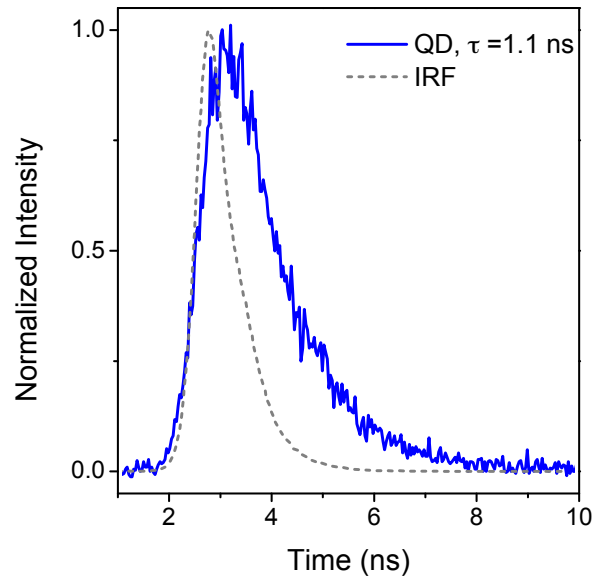

Supplementary Figure 2 – Time-resolved PL of a typical QD transition (blue) and the instrument response function (IRF, black)

## 2. SAW Amplitude dependence

In contrast to our previous experiments on highly strained InGaAs/GaAs QDs [2,4], we do not observe switching between different charge states for the strain-free GaAs/AlGaAs sensor QDs used here. The characteristic oscillations for InGaAs arise from spatio-temporal charge carrier dynamics in the surrounding wetting layer driven by the piezoelectric field of the SAW. For the QDs studied here, the driving piezoelectric field is significantly screened by free carriers being photogenerated in the GaAs regions. In particular, a highly conductive layer can form by accumulating at the AlGaAs-GaAs interface below the QDs induced by the vertical electric field component of the SAW. Therefore, only the mechanical component of the SAW significantly contributes to the spectral modulation [5]. As shown in Reference [6], the deformation potential coupling should exhibit a linear dependence on the acoustic amplitude $A_{SAW} \propto \sqrt{P_{RF}}$. In Supplementary Fig. 3 we summarize the measured spectral modulation of single QDs as a function of $P_{RF}$ for $f_1 = 182.7$MHz (black symbols) and $f_3 = 548.1$ MHz (blue symbols). The data is plotted in double logarithmic representation. Linear fits reveal exponents of $m = 0.58$ and $m = 0.57$ for the two frequencies, close to the expected value of $m = 0.5$ (red line). The green line marks an $m = 1$ power law characteristic for Stark effect tuning. Clearly, our experimental data confirms deformation potential coupling as the underlying tuning mechanism and excludes a dominant contribution of acoustically mediated tuning [6].

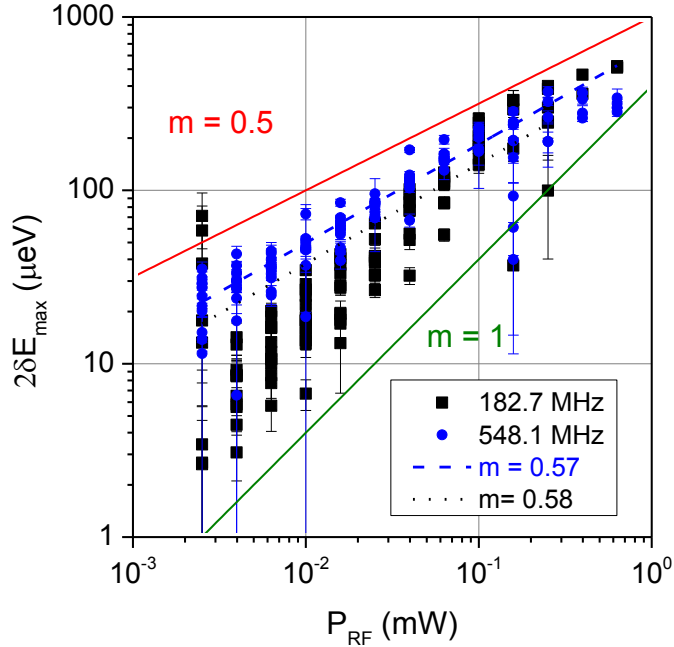

Supplementary Figure 3 – RF-power dependence of spectral modulation bandwidth of single QD emission lines for two frequencies.

### 3. Optomechanical coupling parameter $\gamma_{om}$

In general, the optomechanical coupling parameter defined in Equation 3 of the main manuscript can be rewritten as follows:

$$\gamma_{om} = \left.\frac{\partial(\delta E_{max})}{\partial u_z}\right|_{z=0} = \frac{dE_{gap}}{dp} \cdot \frac{\partial p(z=-152\text{nm})}{\partial u_z(z=0)}. \text{ (SE1)}$$

In this equation $\frac{dE_{gap}}{dp}$ denotes the strength of the DP coupling in hydrostatic approximation. The strength of the DP coupling we used $\frac{dE_{gap}}{dp} = 150\frac{\mu eV}{MPa}$ as reported by Qiang, Pollak and Hickman for $Al_{0.22}Ga_{0.78}As$ [8]. We want to note that this value for the ternary compound is approximately 25% larger than the value for bulk (binary) GaAs reported by Pollak and Cardona [9]. This value is in the sense of optomechanical interactions universal since it only depends on the material system and is independent on the applied mechanical actuation mechanism in the linear (Hooke) regime.

*Evaluation of $\gamma_{om}$ from numerical simulations*

Following (SE1), $\partial p$ is evaluated at the position of the QD at $z$ = -152nm. Here, each Fourier component contributes differently due to the pronounced $z$-dependence of $p$. This partial differential is related to that of the vertical displacement ($\partial u_z$) at the surface ($z = 0$). In our setup we the QD is located in close vicinity to this reference point of the mechanical motion. Thus, the resulting lever arm between the reference point of the mechanical motion and the QD is small compared to NW-based architectures [10,11] for which this reference point is separated by several microns from the optical emitter. We derive $\gamma_{om}$ by first evaluating the calculated $\frac{\partial p(z=-152\text{nm})}{\partial u_z(z=0)}$ for all four elementary NMWFs. In a second step, we obtain $\gamma_{om}$ using (SE1) and $\frac{dE_{gap}}{dp} = 150\,\frac{\mu eV}{\text{MPa}}$. The such obtained values of stated for all elementary NMWFs in the main manuscript.

*Results of analysis*

In this subsection we demonstrate this procedure for the four NWMFs. In Supplementary Figs 4, 5, 6 and 7 we plot the calculated vertical displacement $u_z(z = 0)$ (blue) and $p(z = -152\text{nm})$ (red) in panels a. In panels b the $p(z = -152\text{nm})$ (blue symbols) is correlated to $u_z(z = 0)$. $\frac{\partial p(z=-152\text{nm})}{\partial u_z(z=0)}$ is given by the slope of this curve which is evaluated by a best fit to the data plotted as a red solid line. In panels c $\frac{\partial p(z=-152\text{nm})}{\partial u_z(z=0)}$ and $\gamma_{om}$ obtained from equation (SE2) are evaluated and plotted as a function of $u_z(z = 0)$. For a single frequency sine wave $\gamma_{om}$ is constant as expected. For the square and sawtooth wave, large slopes are observed for $u_z(z = 0) = 0$, i.e at the fast edges of the NMWF. Note that for the sawtooth wave the rising and falling edges are identical, while for the sawtooth wave the corresponding slopes are different as seen in the data. For the δ-pulse maximum tuning occurs at $u_z(z = 0) < 0$. Thus, we evaluate $\gamma_{om}$ at the center of the edges which matches the slope $m$ observed in the data in panel b. The derived values of the tuning slope $m$ and $\gamma_{om}$ are given for each value in panels b and c, respectively.

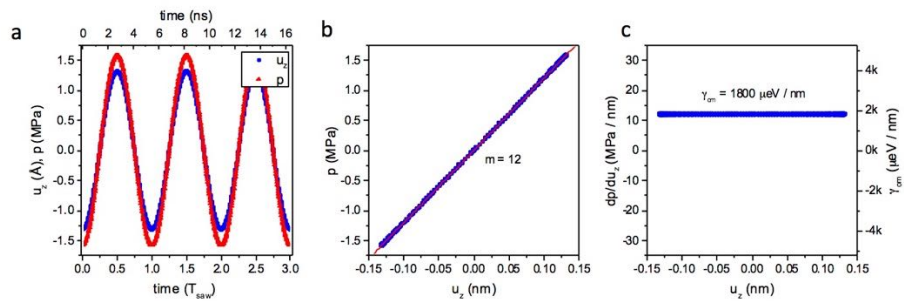

Supplementary Figure 4 – Analysis of $\gamma_{om}$ for sine wave. a – displacement $u_z(z = 0)$ (blue) and hydrostatic pressure $p(z = -152\text{nm})$ (red) as function of time during the fundamental cycle. b – $p(z = -152\text{nm})$ at the QD position as a function of $u_z(z = 0)$. Linear fit to the data (red) and corresponding slope $m$. c – Derivative $\frac{\partial p(z=-152\text{nm})}{\partial u_z(z=0)}$ and corresponding $\gamma_{om}$.

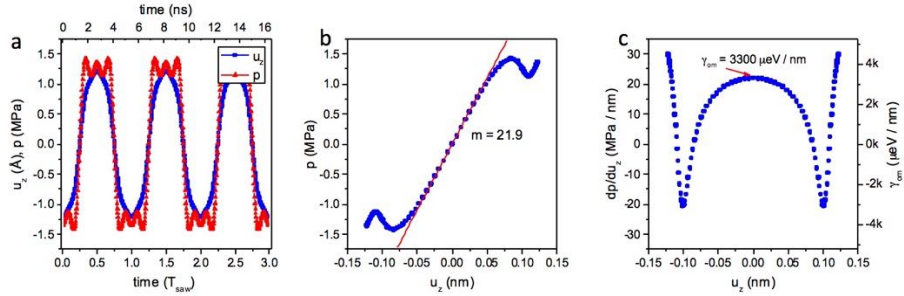

Supplementary Figure 5 – Analysis of $\gamma_{om}$ for square wave. a – displacement $u_z(z=0)$ (blue) and hydrostatic pressure $p(z=-152\text{nm})$ (red) as function of time during the fundamental cycle. b – $p(z=-152\text{nm})$ at the QD position as a function of $u_z(z=0)$. Linear fit to the data (red) and corresponding slope $m$. c – Derivative $\frac{\partial p(z=-152\text{nm})}{\partial u_z(z=0)}$ and corresponding $\gamma_{om}$.

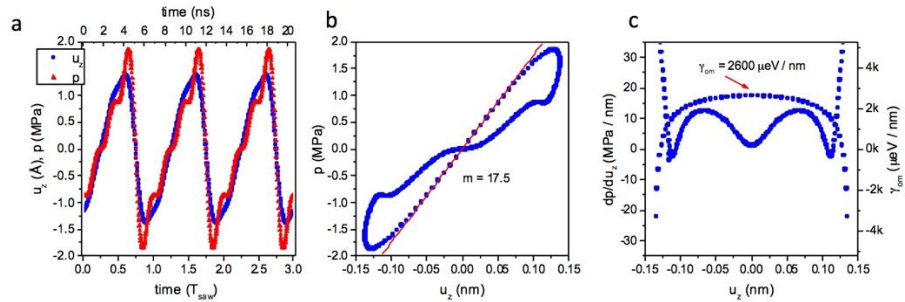

Supplementary Figure 6 – Analysis of $\gamma_{om}$ for sawtooth wave. a – displacement $u_z(z=0)$ (blue) and hydrostatic pressure $p(z=-152\text{nm})$ (red) as function of time during the fundamental cycle. b – $p(z=-152\text{nm})$ at the QD position as a function of $u_z(z=0)$. Linear fit to the data (red) and corresponding slope $m$ for the fast edge of the NMWF. c – Derivative $\frac{\partial p(z=-152\text{nm})}{\partial u_z(z=0)}$ and corresponding $\gamma_{om}$.

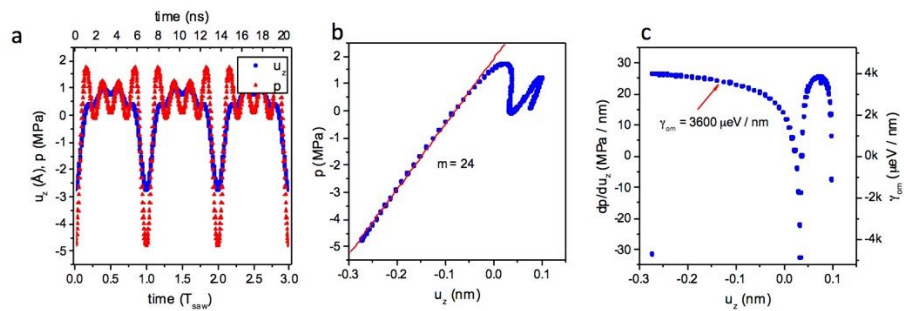

Supplementary Figure 7 – Analysis of $\gamma_{om}$ for δ-pulse. a – displacement $u_z(z=0)$ (blue) and hydrostatic pressure $p(z=-152\text{nm})$ (red) as function of time during the fundamental cycle. b – $p(z=-152\text{nm})$ at the QD position as a function of $u_z(z=0)$. Linear fit to the data (red) and corresponding slope $m$ for the fast edge of the NMWF. c – Derivative $\frac{\partial p(z=-152\text{nm})}{\partial u_z(z=0)}$ and corresponding $\gamma_{om}$. The stated value is extracted at the center of the edges and matches the slope $m$ derived in panel b.

*Comparison to nanowire based architectures*

From the data in Fig. 3e of Yeo and coworkers [10] demonstrating $\gamma_{om} \approx 165 \frac{\mu eV}{nm}$, we derive an enhancement of more than one order of magnitude for our SAW technique. Using the value reported by Montinaro and coworkers [11] of $\gamma_{om} \approx 9.9 \frac{\mu eV}{nm}$, which is smaller than that reported by Yeo et al due to the symmetric shape of their NW. Using the latter value and taking into account for a 20 μm long NW, we would expect from a simple lever arm argument $\gamma_{om} = 9.9 \cdot \frac{20}{0.152} = 1300 \frac{\mu eV}{nm}$ in good agreement with our experimentally derived values. The enhancement of $\gamma_{om}$ by Yeo et al. compared to that by Montinaro et al. can be understood by the conical shape of the etched NWs by Yeo et al. for which the asymmetric mass distribution yields a higher strain at the anchoring point for the same deflection of the NW end.

4. *Fourier analysis procedure*

The Fourier analysis presented in the main paper presented in Figure 4 was performed on the data presented in Figure 3 using a sine-wave basis. We performed a best fit using the following equation

$$\begin{aligned}\delta E(t) = &E_1 \sin[n_1(\omega_1 - \varphi_1)t] \\ &+ E_2 \sin[n_1(\omega_1 - \varphi_2 - \varphi_1)t] \\ &+ E_3 \sin[n_1(\omega_1 - \varphi_3 - \varphi_1)t] \\ &+ E_4 \sin[n_1(\omega_1 - \varphi_4 - \varphi_1)t]\end{aligned} \quad \text{(SE2)}$$

In Equation (SE2) $n_i$ denote the different Fourier components, $E_i$ their corresponding amplitudes and $\varphi_i$ the relative phase offsets with respect to the fundamental ($n_1$=1). Note, that in equation (1) of the main manuscript, the δ-pulse is defined in a cosine-wave basis while this Fourier analysis was performed in a sine-wave basis. Due to the finite temporal resolution, the amplitudes of the different harmonics are more susceptible to error, while the phase information is preserved better. Using the two different basis we are able to demonstrate the excellent agreement between the programmed and measured NMWF in the *finite, non-zero* relative phases.

For the square wave, we expected the non vanishing Fourier coefficients scaling as $1/n_i$ for its odd harmonics ($n_{1,3,5}$). For the sawtooth and δ-pulse both even and odd harmonics contribute. Again, the Fourier coefficients are expected to scale like $1/n_i$ for the sawtooth wave, while for the δ-pulse constant coefficients are expected for all harmonics. Note that the δ-pulse defined in (1) of the main paper consists of a superposition of cosine functions in contrast to the other three waveforms which are defined as superpositions of sine functions. Thus, the expected $\varphi_{2-4}$'s are non-zero for the δ-pulse and zero for the other three waveforms. In Supplementary Fig. 8, we compare the experimental data (symbols) to the result of this fit (green line) which confirms further only minute deviations from the programmed waveform (dashed red line). Details of these small deviations are given in the main text.

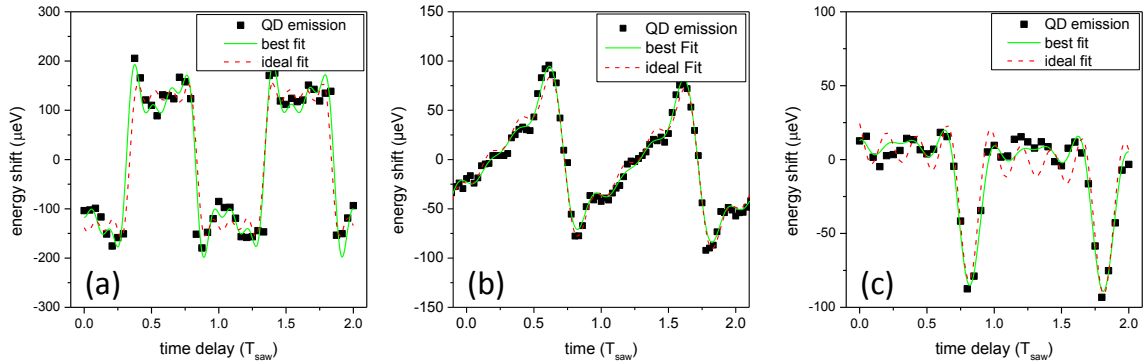

Supplementary Figure 8. – rf-power dependence of spectral modulation bandwidth of single QD emission lines for two frequencies.